\documentclass[aps]{revtex4}
\usepackage{amsmath}
\usepackage{graphicx}
\def\ra{\rangle}
\def\la{\langle}

\def\ra{\rangle}
\def\la{\langle}

\begin{document}
\title{First bromine doped cryogenic implosion at the National Ignition Facility}
\author{A. C. Hayes$^*$, G. Kyrala, M. Gooden, J.B. Wilhelmy, L. Kot, P. Volegov, C. Wilde, B. Haines, Gerard Jungman, R.S. Rundberg, and D.C. Wilson}
\affiliation{Los Alamos National Laboratory, Los Alamos, NM, 87545 USA}
\author{C. Velsko, W. Cassata, E. Henry, C. Yeamans, C. Cerjan, T. Ma, T. D\"oppner, A. Nikroo, O. Hurricane,  D. Callahan, D. Hinkel, D. Schneider, B. Bachmann, and F. Graziani}
\affiliation{Lawrence Livermore National Laboratory, Livermore, CA, 94551 USA} 
\author{K. C. Chen, C. Kong, H. Huang, J. W. Crippen, M. Ratledge, N.G. Rice, and M. P. Farrell}
\affiliation{General Atomics, San Diego, CA,  92121 USA}
\date{\today}

\begin{abstract}

We report  on  the  first  experiment  dedicated  to  the  study  of  nuclear 
reactions on dopants in a cryogenic capsule at the National Ignition Facility (NIF). 
This was accomplished using bromine doping in the inner layers of the CH ablator of a capsule identical to that used in the NIF shot N140520.
The capsule was doped with 3$\times$10$^{16}$ bromine atoms.
The doped capsule shot, N170730, resulted in a DT yield that was 2.6 times lower than the undoped equivalent.
The Radiochemical Analysis of Gaseous Samples (RAGS) system was used to collect 
and detect $^{79}$Kr atoms resulting from energetic deuteron and proton ion reactions on $^{79}$Br. 
RAGS was also used to detect $^{13}$N produced dominantly by knock-on deuteron reactions on the $^{12}$C in the ablator.
High-energy reaction-in-flight neutrons were detected via the $^{209}$Bi(n,4n)$^{206}$Bi reaction, using bismuth activation foils 
located 50 cm outside of the target capsule.
The robustness of the RAGS signals suggest that the use of nuclear reactions on dopants as diagnostics is quite feasible.
\end{abstract}

\maketitle
$^*$ Corresponding author: {\it anna$\_$hayes@lanl.gov}

\section{Introduction}
Fusion ignition has been achieved \cite{ignition}  at the  National Ignition Facility (NIF) using tri-layered cryogenic capsule designs,
and such designs now routinely produce 
inertial confined fusion (ICF) energy gains that
exceed the amount of energy deposited into the central hotspot region \cite{Lawson, Zylstra,Omar, LePape}. These high-yield designs
 open the possibility of studying nuclear reactions under novel plasma conditions.   
The central deuterium-tritium fuel hotspot 
region of cryogenic capsules is surrounded by a cold dense DT fuel layer, and the cold fuel is further surrounded by an outer ablator layer.
Variations on these capsule designs, involving small amounts of a dopant 
in the ablator material, can be used to study nuclear reactions in a burning inertial confinement fusion capsule.
Here we report on the first NIF experiment dedicated to the study of nuclear  reactions on dopants in 
a cryogenic capsule. 
The goal of the experiment was two-fold. First, it was to test the degree to which the fusion yield would be affected by the presence of 
ablator dopants. 
Second, it was designed to measure nuclear reactions on a dopant material 
that is being considered as a mix diagnostic for alternate ICF designs, particularly designs that are opaque to x-rays.
For example, in Double Shell (DS) designs \cite{DS} mix of the high-Z shell into the DT fuel could have an important effect on capsule performance,
 but as yet a robust mix diagnostic for DS designs has not been fully developed.
One possibility is to use the sensitivity of charged-particle radiochemical reactions (Radchem) to mix \cite{betamix}, by measuring Radchem yields on dopants in the  shell of a DS capsules.
The effect of mix on Radchem is complicated because, on one hand, mix causes the dopant material in the ablator (or the shell) to move closer
to the charged-particle fluence born in the DT fuel, while, on the other hand, the mix increases the plasma stopping power  and suppresses these fluences.
For the Double Shell program at NIF, we plan to use reaction-in-flight (RIF) neutrons and  alpha-induced
and knock-on deuteron  Radchem to study mix. For this, the high-Z shell will be doped with both $^{10}$B and $^{79}$Br
and the $^{10}$B($\alpha$,n)$^{13}$N and $^{79}$Br(d,2n)$^{79}$Kr reaction yields measured using the Radiochemical Analysis of Gaseous Samples (RAGS) \cite{Dawn}
facility.
While RIF neutrons are routinely observed on cryogenic shots at NIF, 
demonstrating the feasibility of using $^{79}$Br(d,2n)$^{79}$Kr as a diagnostic is a central goal of the current work. 
 
 RIF spectra are always reduced \cite{hayes2} by the presence of mix, whereas $\alpha$- or $d$-induced reactions are  usually enhanced \cite{betamix}. 
In principle, experimental 14 MeV and down-scattered neutron images could be post-processed to deduce the knock-on ion and alpha particle fluences, if some assumptions are  made about the plasma temperature and density using a combination of experimental data and simulations.
 The  expected RIF and Radchem signals could then be calculated for the  no-mix case  
and compared with the measured signals to deduce the degree of mix taking place.
Current on-going experimental efforts at NIF are aimed at determining the accuracy of such a  procedure.
The present experiment is more motivated by the need for a  proof-of-principle demonstration that charged particle reactions on Br dopants are measurable. 

NIF cryogenic capsules are imploded along a trajectory in pressure-density space that keeps the  
ice layer cold and dense. 
A fraction of the 14 MeV neutrons produced in the hotspot elastically scatter with the D and T ions in the cold fuel,  
 producing a fluence of knock-on (KO) D and T ions.
The KO ions can then undergo nuclear reactions with other ions in the cold fuel or with the ablator material and/or its dopants.
In addition to KO ions, there are sources of energetic protons that can also produce nuclear reactions with the dopant material.
These proton sources include neutron-induced breakup of the D ions and neutron elastic scattering with the hydrogen ions in the CH ablator.
The interaction of the latter source of protons with dopants in high-Z shell designs is unlikely, but is a concern in the present 
experiment and below we estimate how much of our $^{79}$Kr signal was induced by up-scattered ablator hydrogen. 

The  $d+t\rightarrow\alpha+n_{RIF}$ reactions taking place in the cold fuel can be measured
through the spectrum of reaction-in-flight (RIF) neutrons ($n_{RIF}$)  produced with
energies up to 30 MeV \cite{hayes1, hayes2, hayes3}.
The products from KO ion reactions and/or energetic protons with ablator or dopant material can be detected using the
RAGS facility \cite{Dawn} at NIF.
The RAGS facility has demonstrated that 
 $^{13}$N has been detected \cite{cerjan} from the $^{12}$C(d,n)$^{13}$N reaction 
arising from  KO deuteron ions interacting with
the carbon in the CH ablator. 
In the latter work, it was noted that a small faction ($\sim$ 3\%) of $^{13}$N could also be produced by the $(p,n)$ 
reaction on $^{13}$C.
More recently, $^{13}$N was detected \cite{Diego} at RAGS from the $^{10}$B($\alpha$,n)$^{13}$N reaction in a $^{10}$B doped exploding pusher capsule design shot at  NIF. 
The present experiment was designed to use the RAGS facility to measure reactions between KO ions and/or energetic protons and the bromine dopants that 
were loaded into the first inner most few microns of the capsule ablator.

\section{Capsule Design}
The capsule design was based on one of the  first of a series of 
high-foot cryogenic shots to achieve alpha self-heating, shot N140520 \cite{Omar2,Omar3}. 
The ``high-foot" platform at NIF involves a laser pulse-shape that is designed to produce an indirectly driven 3-shock implosion on a relatively high adiabat that helps to control the symmetry and possible instabilities of the implosion.  
The capsule was a so-called T-1 design, meaning that the  capsule involved a 935 $\mu$m inner region filled with DT fuel, surrounded by a 175 $\mu$m thick ablator that involved three layers of silicon doping as a pre-heat shield.
For the present shot, N170730, the ablator   
 was doped with  bromine in order to search for krypton production
via the
$^{79}$Br(d,2n)$^{79}$Kr and $^{79}$Br(p,n)$^{79}$Kr reactions.
RAGS pumps the gas from the NIF chamber after a shot, 
and the collected radioactive gaseous species are assayed.
Several tests of the RAGS collection efficiency have been made using exploding pusher designs \cite{Dawn}.
In general, it is found that noble gases are collected with high efficiency and the collection fraction is not dependent on the experimental setup of
the shot. For example, the collection efficiency is the same with or without a hohlraum. On the other hand, $^{13}$N is collected with
low efficiency and whether its collection from a capsule is affected by issues such as the existence of a hohlraum is, as yet, unknown and is currently under study at NIF.     

\begin{figure}[h]
\includegraphics[width=15 cm]{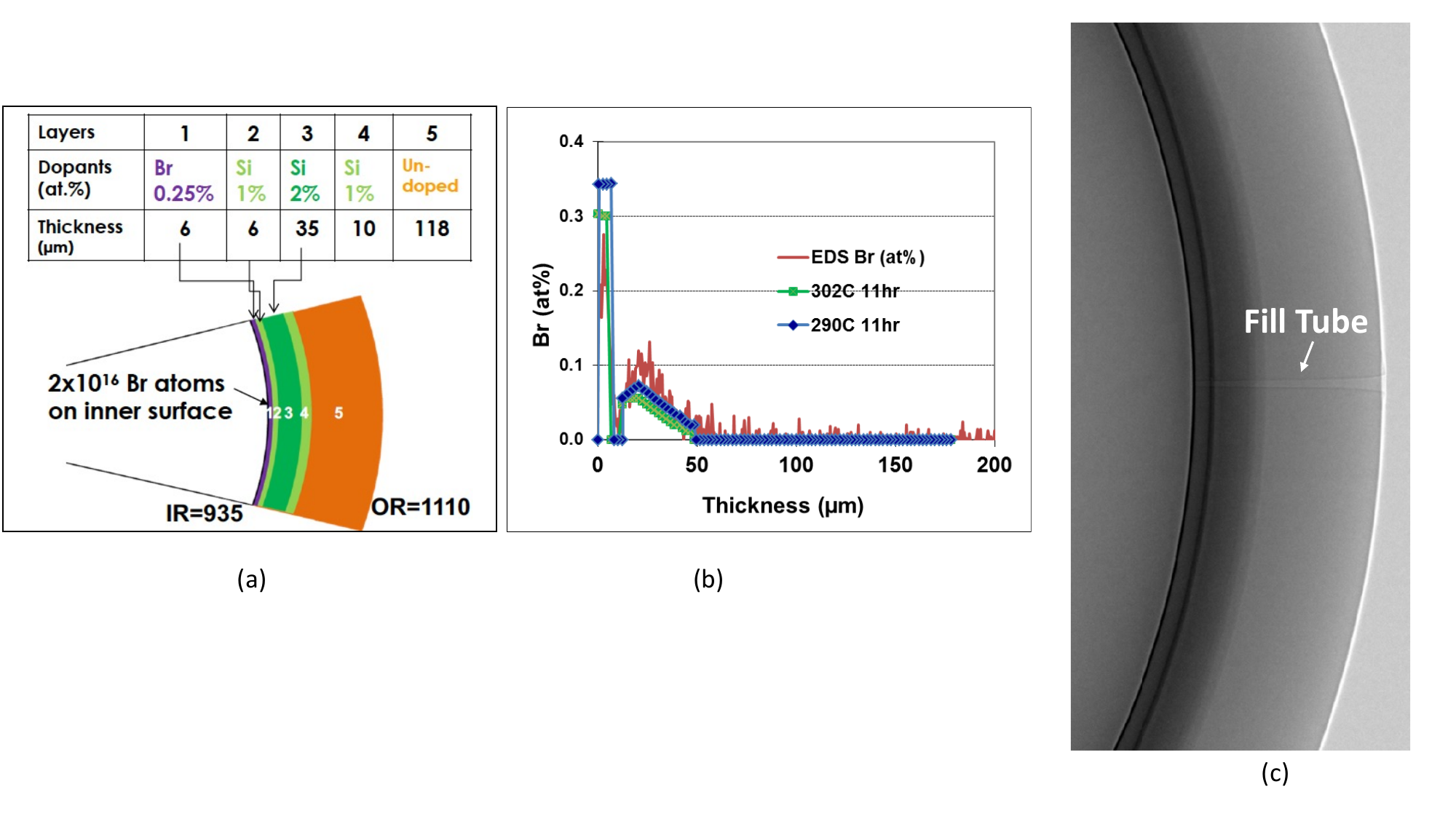}
\caption{ The bromine-doped capsule used in the experiment.
(a)The designed capsule involved a 935 $\mu$m inner region, to be filled with DT fuel, surrounded by a 175 $\mu$m thick ablator.
The CH ablator was designed to involve 5 regions, an inner Br-doped region, followed by 3 Si-doped layers, and an outer undoped CH layer.
(b)The final Br-dopant profile, which involved some leaking of the Br into the first 2 Si-layers. 
The red data in panel (b) shows the Br concentration as measured by energy dispersive spectroscopy (EDS). 
The blue and green data in the same panel show the position of the vaporization and diffusion of the underlying
 soft plastic mandrel through the Si and Br layers at 302$^\circ$ C and 290$^\circ$ C, respectively.  The Br-dopant in the manufactured capsule was present in the inner 50 $\mu$m of the ablator.
(c) The final capsule, shown here at room temperature, involved no visible delamination after drilling for the DT gas fill tube. 
}
\label{capsule}
\end{figure}

The capsule was fabricated by General Atomics (GA). The ablator involved five CH layers. In the original design of the capsule, the
inner most layer was doped with about 3x10$^{16}$ Br atoms, and the next three layers were doped with Si, and the outside layer was undoped CH.
However, the fabrication of the capsule involved several challenges and the bromine dopant in the fabricated capsule went to a depth of 50 $\mu$m.  
Much of the initial added bromine was lost during
the pyrolysis process, but some of the escaping bromine got trapped in the the Si layers. 
An additional complication is that delamination between the Br layer and the adjacent Si layer can occur.
However, the GA team overcame several of these problems and the target used in the presently reported experiment is shown in Fig.\ref{capsule}, along with the original design.

The capsule was driven indirectly using a depleted uranium hohlraum, filled with 1.6 mg/cc $^4$He gas and cooled at shot time. The hohlraum design is described in detail in \cite{doppner}.
\section{The physics governing reactions on dopant material}
\subsection{Effect of Bromine Dopant on the Shot Yield}
The neutron yield of shot N170730 was $2.9\pm 0.1\times 10^{15}$, to be compared to $7.6\pm 0.3\times 10^{15}$ for the analogous undoped shot N140520.
The hotspot ion temperature was also lower: $4.36 \pm 0.12$ keV for N170730 and $5.6\pm0.13$ keV for N140520.
The  temperature difference is consistent with the difference in the yields for the two shots, i.e., the ratio of the corresponding reactivities ($\langle \overline{\sigma v}\rangle$) \cite{Caughlan} is 2.47, compared to the ratio of the yields, 2.6.
Our pre-shot simulations suggested that a reduction in the yield could  occur, mostly because 
of an increase in the  entropy of the cold fuel induced by mixing of some of the bromine doped layer into this region. 
In choosing the dopant level, we aimed to achieve a balance between having enough dopant target atoms for a robust signal at RAGS
and keeping the dopant level low enough to avoid a detrimental effect on the DT yield. 
Two independent 1-D pre-shot simulations were carried out to obtain estimates for the possible effect of the dopant on capsule performance. 
In the first, we simply assumed, without justification, 
that all of the the ablator material containing bromine dopants was uniformly mixed into the cold fuel. 
This caused the capsule to implode on a higher adiabat and
resulted in a factor of three reduction in the yield. In the second, we ran a 1-D xRAGE simulation using the BHR
mix model \cite{BHR}, with the mixing length parameter $S_0$ set to 500 nm.  This resulted in non-uniform mixing of some of the 
 ablator material into the cold fuel, and again caused the capsule to implode on a higher adiabat and the predicted yield was a factor of
 2.3 lower than the equivalent  clean calculation.
Whether the existence of the bromine dopant in itself affected the degree of mix taking place is difficult to judge from
this simple  1-D simulation and would require a more in-depth 2-D study. But the 1-D simulations suggest that 
the increase in the implosion adiabat led to a less compressed and cooler hotspot.
Based on these pre-shot simulations we decided that a dopant level of $3\times10^{16}$ bromine atoms was  tolerable.
The pre-shot predictions were in good agreement with the observed yield of N170730, relative to the undoped shot.
But we cannot draw any strong conclusions about the origins of the reduction in yield when bromine dopants were added to the capsule. 
Other factors could have played a role, including possible non-identical laser drive, etc. 

The bangtime for the N170730 doped shot, as measured by gamma reaction history, was 15.87$\pm$0.05 nsec.
The measured radius of the hotspot was P0= 31.7$\pm$2.9 $\mu$m and P0=34$\pm$1.3 $\mu$m from neutron imaging and 
x-ray imaging, respectively.  The neutron imaging P2/P0 ratio was -10\%. By comparison, the neutron imaging size parameters
for N140520 were P0=28 $\mu$m and a P2/P0=-25\%. 
The down-scattered ratio (DSR) for the cold fuel, which is a direct measure of the cold fuel areal density, for the doped 
shot N170730 was only slightly lower (4.0$\pm$0.19) than that for the undoped shot N140520 (4.32$\pm$0.18), so that it is not clear that the compression of the cold fuel was significantly changed by the bromine dopant. 
The neutron imaging data for the shot in the equatorial (90$^\circ$, 315$^\circ$) and polar (0$^\circ$, 0$^\circ$) directions allowed a 
reconstruction of the volume neutron production distribution, Fig. 2,  using the techniques described in \cite{Volegov}.
In the polar direction, neutron imaging saw a radius M0 of 32 $\mu$m and an M2/M0 ratio of +6\%, Fig. 2.
The P0 and P2 values are derived from the Legendre polynomial
fit to the 17\% contour in the equatorial direction, 
while M0 and M2 values are derived in the same way, but in the polar direction.
\begin{figure}[h]
\includegraphics[width=19 cm]{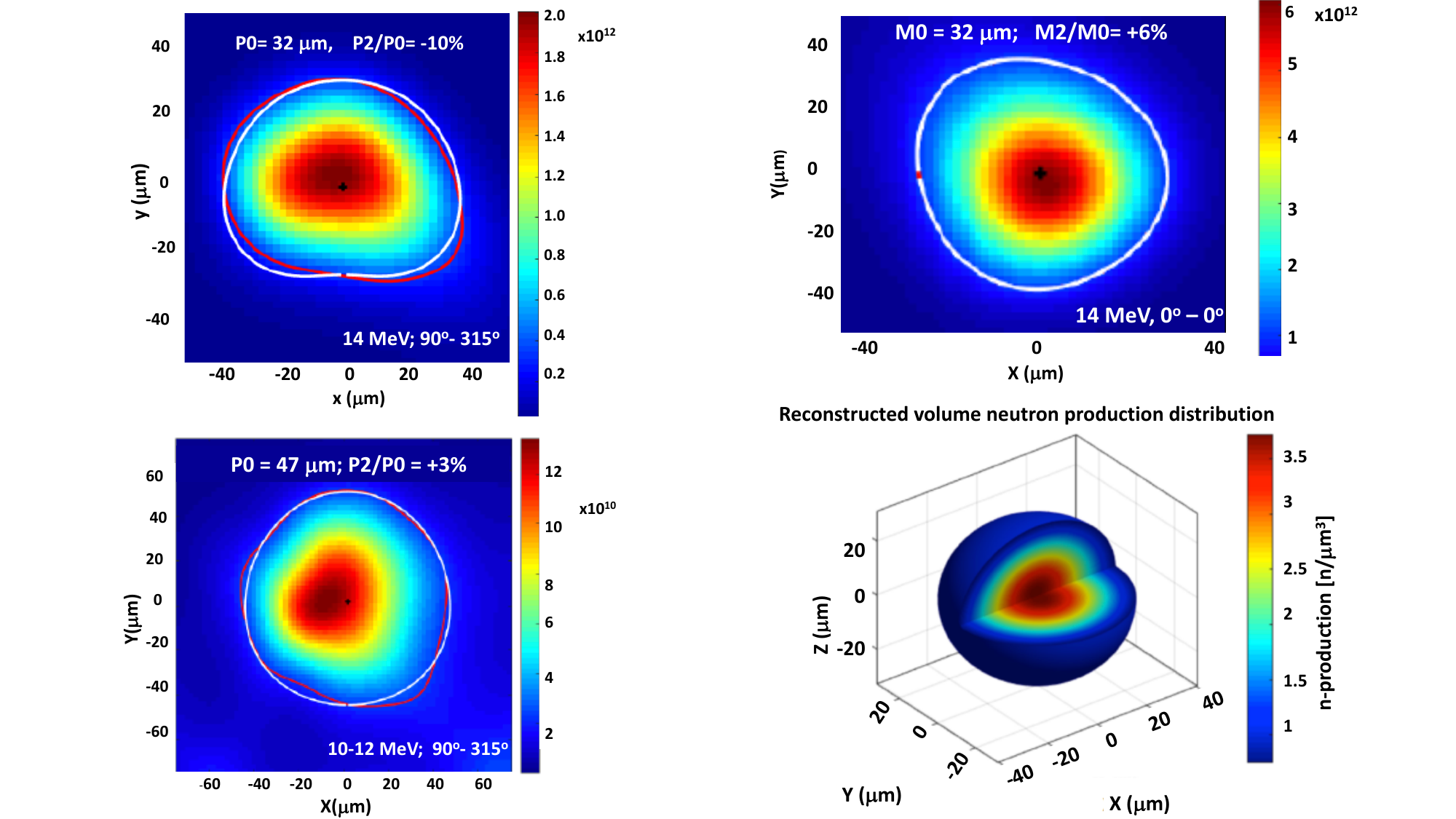}
\caption{ The neutron images for the burning bromine-doped shot N170730. 
The upper left panel is an equatorial image of the primary 14 MeV neutrons produced in the hotspot and suggests a burning region of radius about 32 $\mu$m.
The lower left panel is an equatorial image of the secondary down-scattered (10-12 MeV) neutrons, which provides an assessment of the size of the dense cold fuel.
The upper right image is taken from the polar direction, which shows a hotspot region similar in size, but different in shape, than that seen in the equatorial direction. 
The P0 and P2 values are derived from the Legendre polynomial fit to the 17\% contour in the equatorial direction, 
while M0 and M2 values are derived in the same way, but in the polar direction. 
The lower right panel is a 3-D reconstruction of the volumetric neutron production distribution.}
\label{NIS}
\end{figure}

\subsection{Energetic (MeV)  ion induced reactions}
\begin{figure}[h]
\includegraphics[width=8 cm]{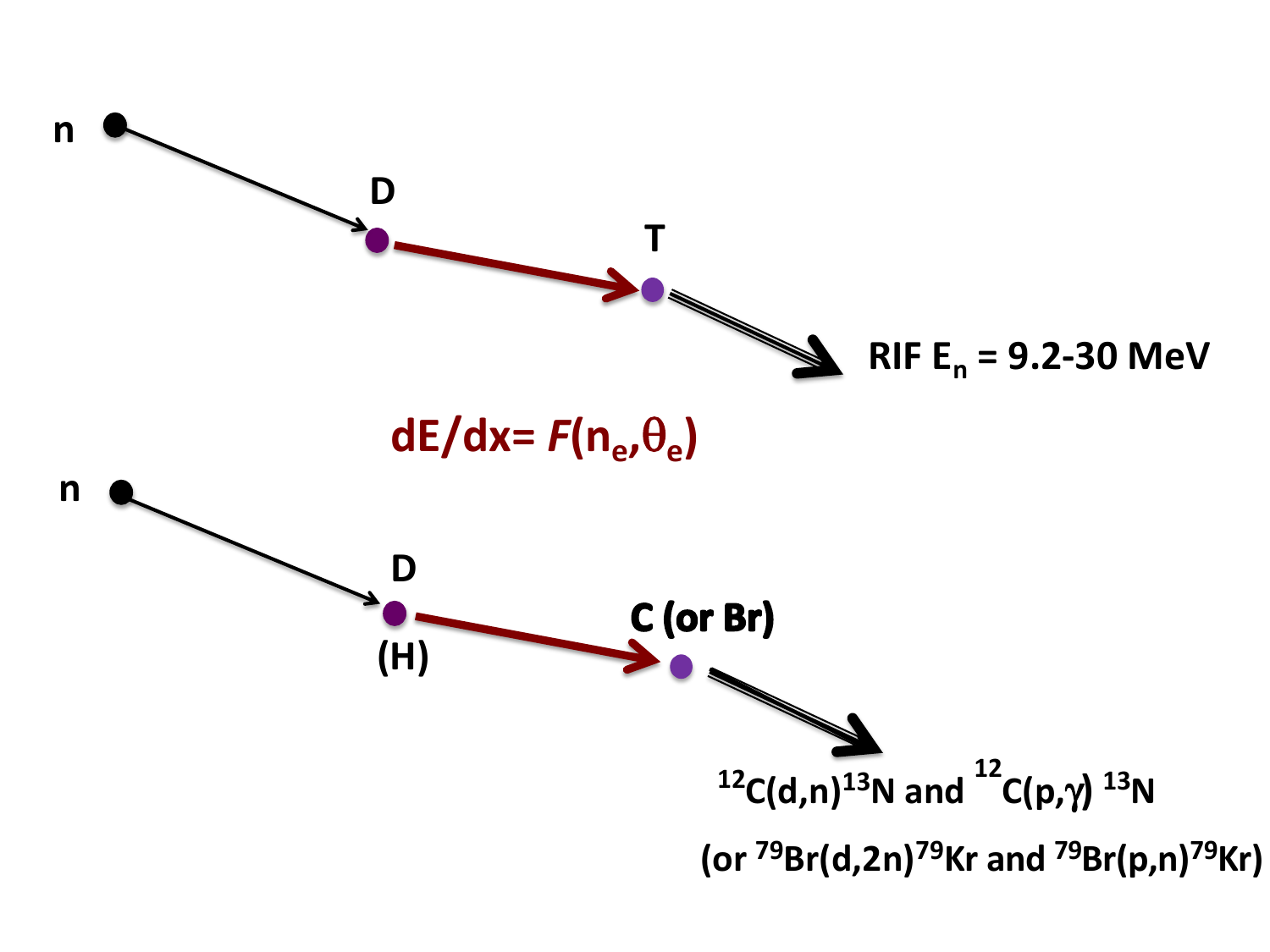}
\caption{ The reactions of interest involve a 14 MeV neutron first producing an energetic P, D or T 
 ion by elastic scattering or by deuteron breakup up,
and the energetic ion then interacting with either a cold fuel ion, ablator material or dopant material.   
Some of the energetic P ion are produced by neutron elastic scattering of the hydrogen in the CH ablator.
As the energetic ion traverses the capsule it looses energy through the plasma stopping power, and the fluence of such ions is inversely proportional to the stopping power.
}
\label{Reactions}
\end{figure}

Our primary interest in this experiment was to measure the yields of three different energetic ion-induced reactions,
namely,  proton and deuteron reactions on Br and C, and DT reactions in-flight. 
The reactions are all analogous, 
and involve an energetic P, D or T  ion first being produced by a 14 MeV neutron interaction and then undergoing a nuclear interaction before loosing all of its kinetic energy, Fig. \ref{Reactions}. We note that the P ions are produced by either deuteron breakup or elastic scattering, while the D and T ions  are produced by elastic scattering. 
The diagnostic interest in these reactions is that these as energetic ions
traverse the capsule they loose energy through the plasma stopping power, and the 
their fluence is inversely proportional
to the stopping power.  
\section{Experimental Radiochemistry Signals}
\subsection{$^{13}$N production}
Immediately after the shot, the NIF chamber was pumped out and the gases collected by the RAGS facility.
The $^{13}$N was collected into the abort tank of RAGS and assayed by counting its decay via positron emission. 
From tests of the RAGS system, we have evidence that a sizable fraction of $^{13}$N in the target chamber can undergo 
chemical reactions and  not be transported to the RAGS system.
We tested the collection and detector efficiency for $^{13}$N using three separate shots, 
in which a boron-nitrate (BN) postage stamp was attached to the outside of the hohlraum of an indirectly driven
exploding pusher. 
In these cases, the $^{13}$N is made via the $^{14}$N(n,2n)$^{13}$N reaction.
The BN postage stamp tests were found  to agree within 5\%.
It is possible that the rate of nitrogen chemical reactions taking place in the NIF chamber is dependent on the configuration 
of the target package, i.e., whether the $^{13}$N produced in the capsule or in the hohlraum.
Understanding these issues remains a primary objective of our current radiochemistry program at NIF.
For this reason, in reporting the current data, we 
increase the uncertainty on our $^{13}$N diagnostic to 20\%, 
and quote 
number of $^{13}$N atoms produced in the Br-doped shot N170730 as 4.80$\pm$0.82$\times$10$^8$ atoms.    
The current radiochemistry  program at NIF, which is focused on the Double Shell campaign, has set a goal of determining radchem inventories at RAGS to better than 10\%. 

\subsection{$^{79}$Kr production}
\begin{figure}[h]
\includegraphics[width=15 cm]{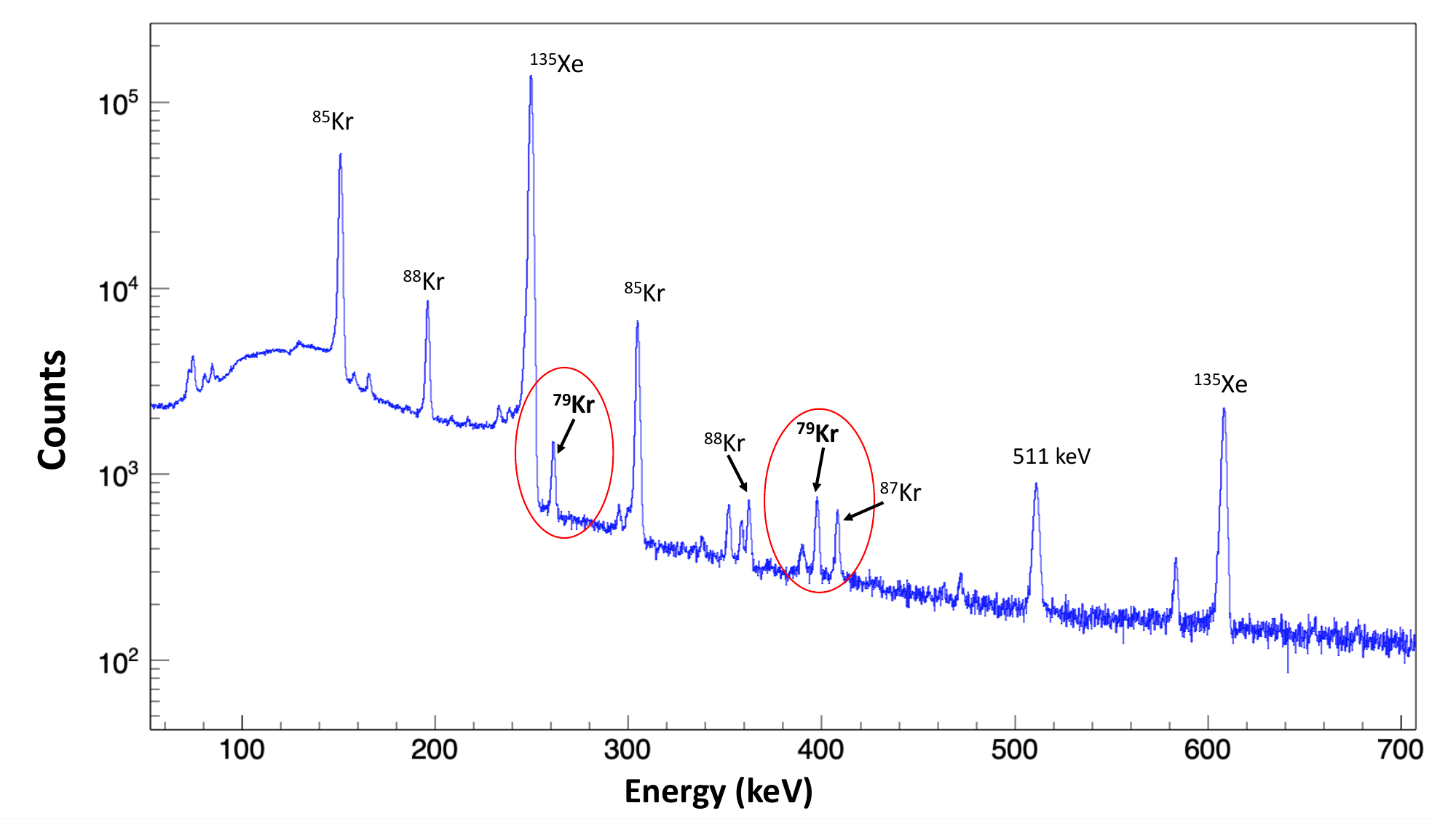}
\caption{ The $^{79}$Kr signal observed by high-resolution $\gamma$-ray detection from the gaseous sample collected by RAGS following the N170730 shot. All of the prominent $\gamma$-rays from the decay of $^{79}$Kr were detected and in the correct expected ratio.
The $^{85}$Kr, $^{88}$Kr and $^{135}$Xe radioisotopes all arise from neutron fissioning of the depleted uranium (DU) hohlraum. $^{79}$Kr is not a fission product, and could only have been produced by proton and deuteron reactions on $^{79}$Br.
}
\label{RAGS}
\end{figure}

We observed a very clear signal for the production of $^{79}$Kr in our assay of the gas collected at RAGS, and all of the prominent
$\gamma$-rays from the decay of $^{79}$Kr were detected and in the correct expected ratio, Fig \ref{RAGS}. 
The measured $^{79}$Kr signal corresponds to a total of 2.86$\times$10$^7\pm$3.4\%  atoms being produced in the target capsule.
Two independent $\gamma$-ray detectors were used to assay the $^{79}$Kr at RAGS and they agree within the quoted error. 
In addition, the collection efficiency of the noble gases of krypton are well characterized at RAGS, so that the uncertainty on the capsule production
of krypton is considerably lower than for $^{13}$N.
We also observed $^{85}$Kr,
$^{88}$Kr and $^{135}$Xe, which were produced in the neutron-induced fission of the depleted uranium (DU) hohlraum. 
$^{79}$Kr is not a fission
product, and could only have been produced by proton and deuteron reactions on $^{79}$Br

\subsection{Estimate of the fraction of $^{79}$Kr atoms produced from KO deuterons}
A main goal of the experiment was to demonstrate the feasibility of using the RAGS facility to detect  $^{79}$Kr produced in the $^{79}$B(d,2n)$^{79}$Kr reaction by knock-on deuterons ions.
But in the present capsule design, $^{79}$Kr could also have been produced in the $^{79}$B(p,n)$^{79}$Kr reaction by elastically scattered hydrogen atoms in the ablator, and we in this section we examine the arguments supporting a non-negligible fraction of the krypton coming from deuteron reactions. 

\begin{figure}
\includegraphics[width=8 cm]{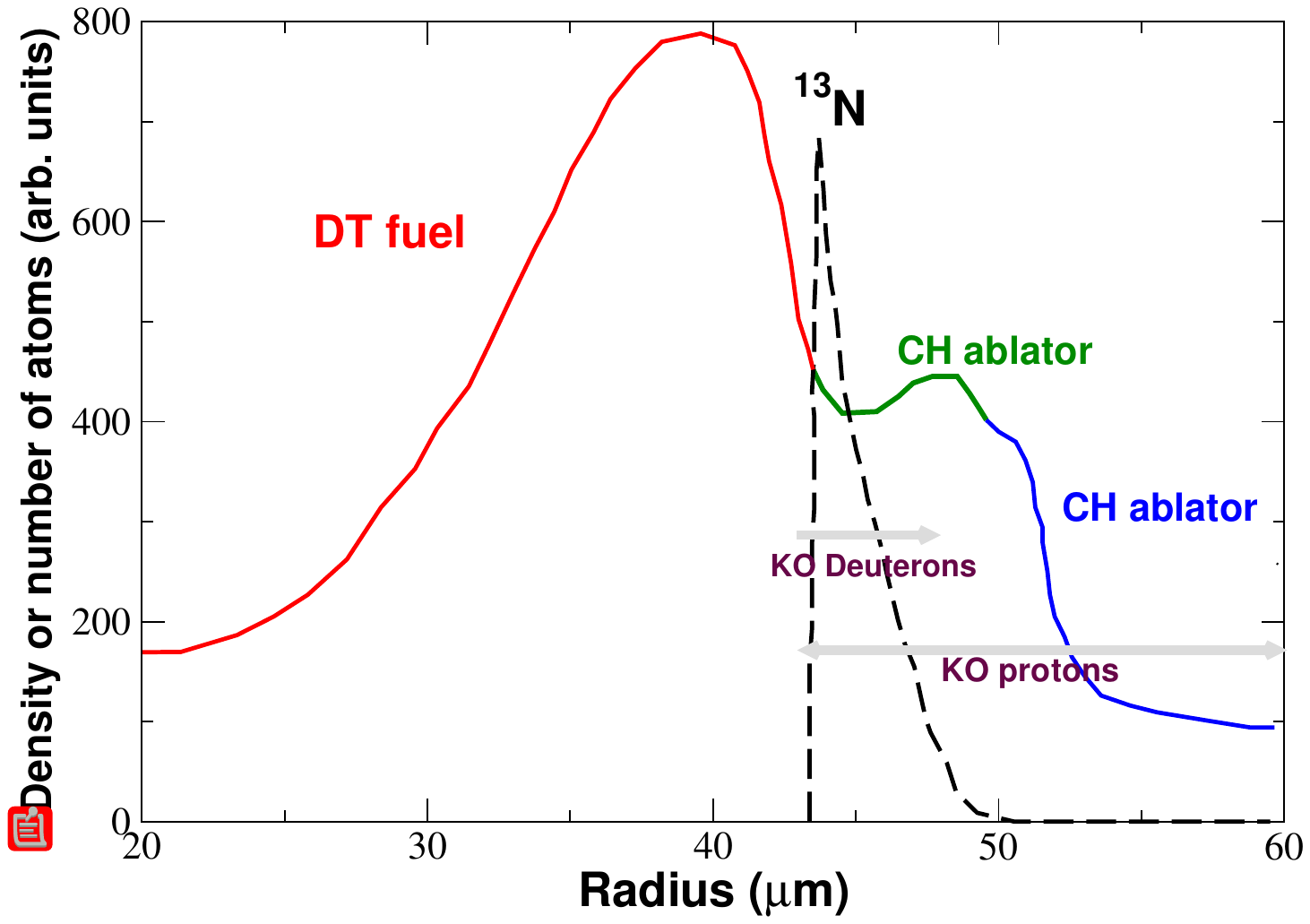}
\caption{The density of the DT fuel and ablator material and the production of $^{13}$N, as a function of the capsule radius, as predicted in a non-mix HYDRA\cite{hydra2, hydra} simulation. 
In this graph, the ablator is divided into two regions, the region in which $^{13}$N was produced (green) and the region in which no $^{13}$N was produced (blue). The dashed black curve shows the $^{13}$N production probability distribution within the 'green' region.
The short range of the KO deuterons in the ablator ($\sim 5\mu$m) determines the size of these regions. 
This two region division of the ablator material applies equally to the deuteron-induced production of $^{79}$Kr and we use the calculated ratio of the $^{79}$Br(d,2n)/$^{12}$C(d,n) production to estimate how much of the 
measured $^{79}$Kr were induced by KO deuterons, as opposed to protons. 
The gray arrows show the region of the ablator in which KO deuterons or KO protons are predicted to be undergoing reactions.
}
\label{13N}
\end{figure}

\begin{figure}
\includegraphics[width=8 cm]{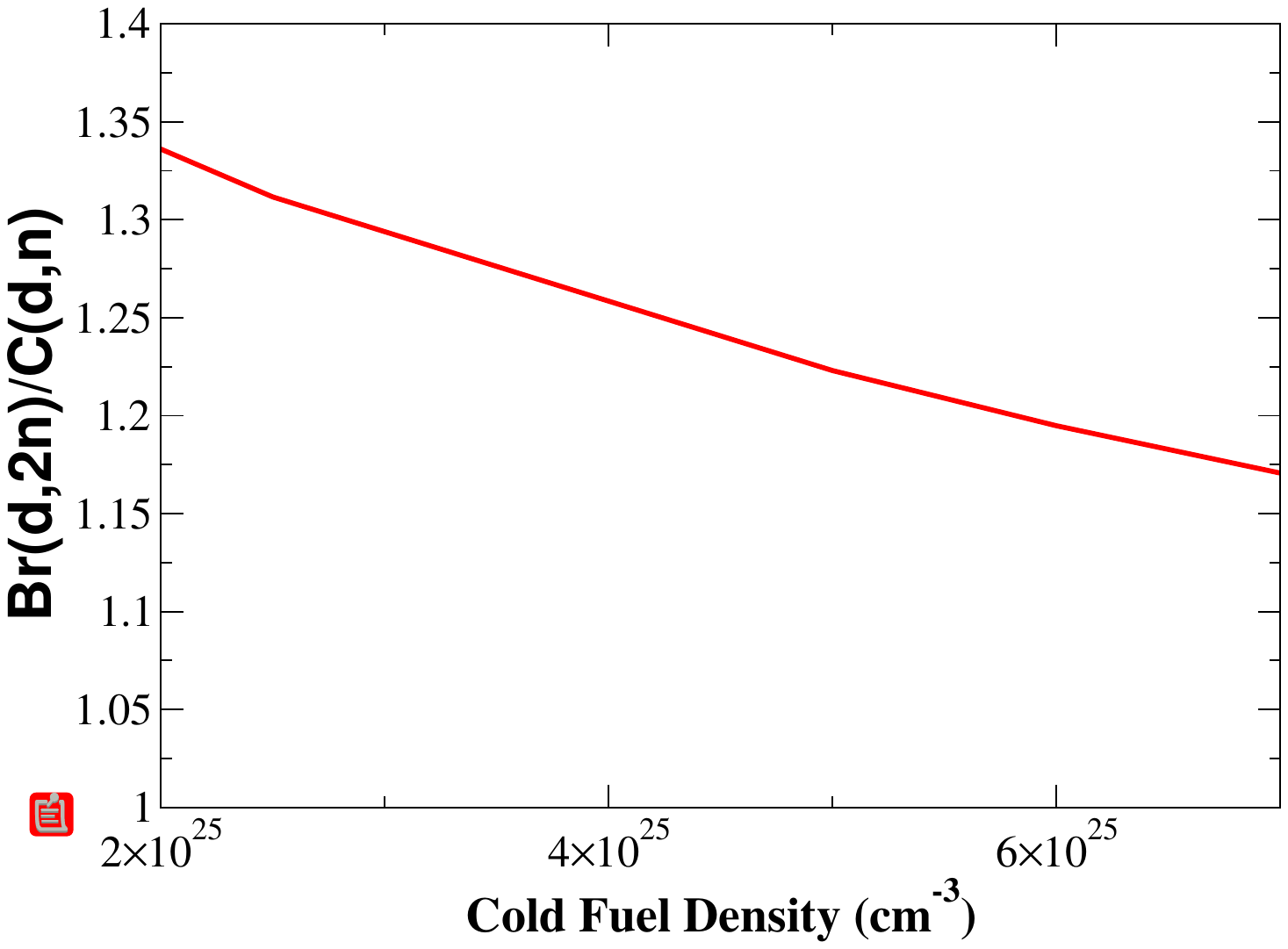}
\caption{The predicted change in the 
ratio of the $^{79}$Br(d,2n)/$^{12}$C(d,n) reaction yields as the shape and magnitude of the KO deuteron fluence is varied through a change in the cold fuel density, 
and assuming an electron temperature of 0.2 keV. 
The ratio only varies by about 14\% over this range of plasma conditions. Varying the temperature within the range expected for the cold fuel has an even smaller effect on this ratio.
Because the deuteron-induced $^{13}$N and $^{79}$Br are produced in the same region of the capsule, we can use the average value of this ratio 
and the measured $^{13}$N production to estimate the amount of $^{79}$Kr atoms produced solely from KO deuterons.  
We find that  7$\times$10$^6$ $^{79}$Kr atoms (or 25\% of the total) were produced from KO deuteron reactions.
}
\label{RATIO}
\end{figure}

Form our simulations, we find that the range of the KO deuterons  in the ablator material is quite short, about 5 $\mu$m. 
This means that all of the deuteron reactions on bromine and carbon most likely took place in the same region of the capsule, 
being either in the first layer of the ablator or in a mixed ablator/fuel region. 
A 2-D prediction for where the $^{13}$N was produced, from a no-mix calculation using a HYDRA \cite{hydra2, hydra} simulation coupled to 
our radiochemistry post-processor  is shown in Fig. \ref{13N}. As can be seen, all of the 
$^{13}$N is predicted within the first 5-6 $\mu$m of the ablator. This is the same region in which the $^{79}$B(d,2n)$^{79}$Kr reactions are predicted to occur.
On the other hand, $^{79}$Br(p,n)$^{79}$Br reactions are predicted to occur in all regions of the ablator containing a finite amount of Br-doping.
To estimate how much of the measured $^{79}$Kr arose solely from the  $^{79}$B(d,2n)$^{79}$Kr reaction, we calculated the $^{79}$Kr/$^{13}$N ratio for 
different reasonable assumptions for the KO deuteron fluence. For this, we varied the cold fuel stopping power by varying the plasma 
density and the temperature.
Fig. \ref{RATIO} shows the variation of the $^{79}$Br(d,2n)$^{79}$Kr/$^{12}$C(d,n)$^{13}$N production ratio as a function of cold fuel density for
cold fuel temperature $\theta^{\rm CF}_e$=0.2 keV.
As can be seen, the ratio only changes by 14\% over the broad density range 2$\times$10$^{25}$ - 7$\times$10$^{25}$ e/cc, with the average ratio being 1.25.
From this and the measured $^{13}$N production, 
 we estimate that about 25\% of the $^{79}$Kr produced in the shot N170730 came from KO deuterons. The other 75\% must have come from 
$^{79}$Br(p,n)$^{79}$Kr reactions.   
 
\subsection{The reaction-in-flight neutrons}
We measured reaction-in-flight neutron 
using neutron activation of a Bi foil that was placed  50 cm outside of the capsule target.  
The $^{209}$Bi(n,4n)$^{206}$Bi reaction 
has
a threshold of E$_n$ = 22.5 MeV and samples an average neutron energy $\la E_n\ra\sim$ 25 MeV.
To measure the RIF signature product $^{206}$Bi we used clover detectors, described in ref. \cite{hayes1}.
Our measured $^{206}$Bi signal is 2344$\pm$376 counts.

\section{summary}
The Br-doped cryogenic shot N170730 involved 3$\times$10$^{16}$ Br atoms in the CH ablator. 
The capsule design was identical to an earlier NIF shot
N140520. 
Table 1 summarizes the measurements made in the present shot, N170730, with comparisons to the identical undoped shot, N140520, where appropriate.
\begin{table}
\begin{tabular}{|c|c|c|c|c|c|c|c|c|c|c}\hline
&Yield&P0&P0/P2&$\theta_{HS}$(keV)&DSR&$^{13}$N&$^{79}$Kr total&$^{79}$Br(d,2n)$^{79}$Kr&$^{206}$Bi\\\hline
N170730&2.9$\pm$0.1$\times$10$^{15}$&32&-10\%&4.36$\pm$0.12&4.0$\pm$0.19&4.8$\pm$0.82$\times$10$^8$&2.86$\times$10$^7$&7$\times$ 10$^6$&$2344\pm376$\\
N140520&7.6$\pm$0.3$\times$10$^{15}$&28&-25\%&5.6$\pm$0.13&4.32$\pm$0.18&-&-&-&-\\\hline
\end{tabular}
\caption{The performance, Radchem and RIF signals for the Br doped shot N170730.}
\end{table}
The yield of the doped capsule was suppressed by a factor of 2.6 relative to the undoped capsule.
Both $^{79}$Kr and $^{13}$N resulting from KO ion interactions with the ablator were detected at RAGS.
The $^{13}$N is dominantly produced from KO deuterons, whereas the $^{79}$Kr can be made with KO deuteron and energetic proton ion induced reactions.
We estimate that about 25\% of the $^{79}$Kr was produced solely by KO deuterons.
We also observed reaction-in-flight neutrons using the $^{209}$Bi(n,4n) reaction. The resulting $^{206}$Bi signal was robust. 

The shot N170730 was successful and it showed that doped cryogenic capsules at NIF can be used to measure charged-particle
 reactions  under plasma
conditions suitable for the production of copious fluences KO ions, especially if the products of such reactions are gaseous and collectible by the RAGS facility.  
N170730 was the first inertial confinement fusion shot in which $^{79}$Kr, produced in $(d,2n)$ and $(p,n)$ reactions, was measured.
This result is particularly encouraging and significant for the DS program, where the high-Z shell renders the hotspot 
 opaque to x-rays.
Current simulations \cite{hayes-SNL} predict the production of $^{79}$Kr 
from charged-particle reactions on bromine dopants in the shell of DS capsules provide sensitive diagnostics of mixing of the  shell into the DT fuel.
In addition, the $^{79}$Br produced in a DS capsule has the advantage of not having any source of energetic protons from the
 ablator material contributing to the diagnostic. 
\subsection{Data availability}
All data appearing in this manuscript are available by contacting Anna Hayes at anna$_\_$hayes@lanl.gov. 
\subsection{Acknowledgments}
This work was performed under the auspices of the U.S. Department of Energy NNSA by Los Alamos National Laboratory under Contract No. DE-AC52-06NA25396 and by Lawrence Livermore National Laboratory under Contract No. DE-AC52-07NA27344.

\end{document}